\documentclass[conference]{IEEEtran}
\IEEEoverridecommandlockouts
\usepackage{cite}
\usepackage{amsmath,amssymb,amsfonts}
\usepackage{algorithmic}
\usepackage{graphicx}
\usepackage{textcomp}
\usepackage{xcolor}
\usepackage{multirow}
\usepackage{soul}
\usepackage[caption=false]{subfig}

\def\BibTeX{{\rm B\kern-.05em{\sc i\kern-.025em b}\kern-.08em
    T\kern-.1667em\lower.7ex\hbox{E}\kern-.125emX}}

\makeatletter
\renewcommand\p@subfigure{\thefigure~(}
\makeatother

\newcommand{\First}{\textcolor{red}{\textbf{$^1$}}}
\newcommand{\Second}{\textcolor{blue}{\textbf{$^2$}}}

\begin{document}

\title{Image Data Hiding in Neural Compressed Latent Representations
\\
\thanks{The authors would like to thank the NSTC of Taiwan and CITI SINICA for supporting this research under the grant numbers 111-2221-E-002-134-MY3 and Sinica 3012-C3447.}
}

\author{
\IEEEauthorblockN{Chen-Hsiu Huang and Ja-Ling Wu}
\IEEEauthorblockA{\textit{Dept. of Computer Science and Information Engineering} \\
\textit{National Taiwan University}, Taipei, Taiwan \\
\{chenhsiu48,wjl\}@cmlab.csie.ntu.edu.tw
}}

\maketitle

\begin{abstract}
We propose an end-to-end learned image data hiding framework that embeds and extracts secrets in the latent representations of a generic neural compressor. By leveraging a perceptual loss function in conjunction with our proposed message encoder and decoder, our approach simultaneously achieves high image quality and high bit accuracy. Compared to existing techniques, our framework offers superior image secrecy and competitive watermarking robustness in the compressed domain while accelerating the embedding speed by over 50 times. These results demonstrate the potential of combining data hiding techniques and neural compression and offer new insights into developing neural compression techniques and their applications.
\end{abstract}


\section{Introduction}

Image steganography or watermarking hides secrets in a cover image to form a container image for communication or proof of ownership. Steganography focuses on the container image's secrecy and message capacity, while watermarking techniques must be robust to various attacks. Traditional methods worked on hiding and extracting secrets in either the spatial domain \cite{pevny2010using,holub2014universal} or the frequency domain \cite{fridrich2007statistically,bi2007robust}. Modern techniques \cite{baluja2017hiding,zhang2020udh,zhu2018hidden,tancik2020stegastamp} use deep neural networks (DNNs) and adversarial training to end-to-end learn an encoder/decoder pair that embeds and extracts the secrets with robustness against noise attacks. These data hiding techniques are highly relevant to image compression codecs and DNN-transformed latent representations.

Neural compression \cite{yang2022introduction}, the end-to-end learned image compression method \cite{balle2018variational,minnen2018joint,cheng2020learned,chen2021end}, has been actively developed in recent years and has proven to outperform traditional expert-designed image codecs. Although the model complexity and performance issues remain the challenges for neural compression to be widely adopted, international standards such as JPEG AI \cite{ascenso2021white} and MPEG VCM (Video Coding for Machines) \cite{evidence2020vcm} have initiated to bridge data compression and computer vision together for both human and machine vision. Choi et al. \cite{choi2022scalable} have proposed scalable image coding frameworks based on well-developed neural compressors to achieve up to 80\% bitrate savings on machine vision tasks.

Data compression itself may not justify the necessity to replace handcrafted image/video codecs with learned approaches. However, the intersection of data compression and multimedia applications may offer us a different perspective. In this work, we propose an end-to-end learned image data hiding framework that embeds and extracts secrets in the latent representations of a neural compressor. 





\section{Related Works}

\subsection{Learned Image Compression}

DNNs have opened new opportunities to rebuild data compression as an end-to-end learning process. While several approaches existed in the literature, Balle et al. \cite{balle2018variational} proposed a highly successful image codec that out-performed JPEG 
and JPEG 2000 
in PSNR and SSIM metrics. Minnen et al. \cite{minnen2018joint} then used a joint autoregressive and hierarchical prior model to achieve even higher coding efficiency than the HEVC \cite{lainema2016hevc} codec. More recently, Cheng et al. \cite{cheng2020learned} developed techniques that achieve comparable performance to the latest coding standard VVC \cite{ohm2018versatile}. There are now several excellent survey and introduction papers \cite{ma2019image,mishra2022deep,yang2022introduction} summarizing this wave of end-to-end compression advances.

\subsection{DNN-based Steganography and Watermarking}

Deep steganography methods like DeepStega \cite{baluja2017hiding} and UDH \cite{zhang2020udh} defined a new task to hide one or more images into a cover image with a DNN. Unlike traditional methods that require a perfect restoration of secret messages, Deep steganography methods minimize the distortion between the retrieved and the original secret images. Thus, the message is securely delivered because the authentic and recovered secret images are visually indistinguishable. Lu et al. \cite{lu2021large} recently advanced deep steganography with a higher capacity of up to three or more secret images.

Zhu et al. \cite{zhu2018hidden} proposed the HiDDeN model that embeds the raw bits and extracts the secret message with a low bit error rate using a DNN. With the generative model adversarial trained against the noise layers, the HiDDeN model can achieve the purposes of steganography and digital watermarking with the same network architecture. Perhaps inspired by HiDDeN, many researchers have proposed similar adversarial network methods for steganography, such as StegaStamp \cite{tancik2020stegastamp}, Hinet \cite{jing2021hinet}, and watermarking \cite{luo2020distortion,luo2021dvmark,wengrowski2019light}.

\section{Proposed Method}

\begin{figure}[!t]
\centering
\includegraphics[width=0.8\columnwidth]{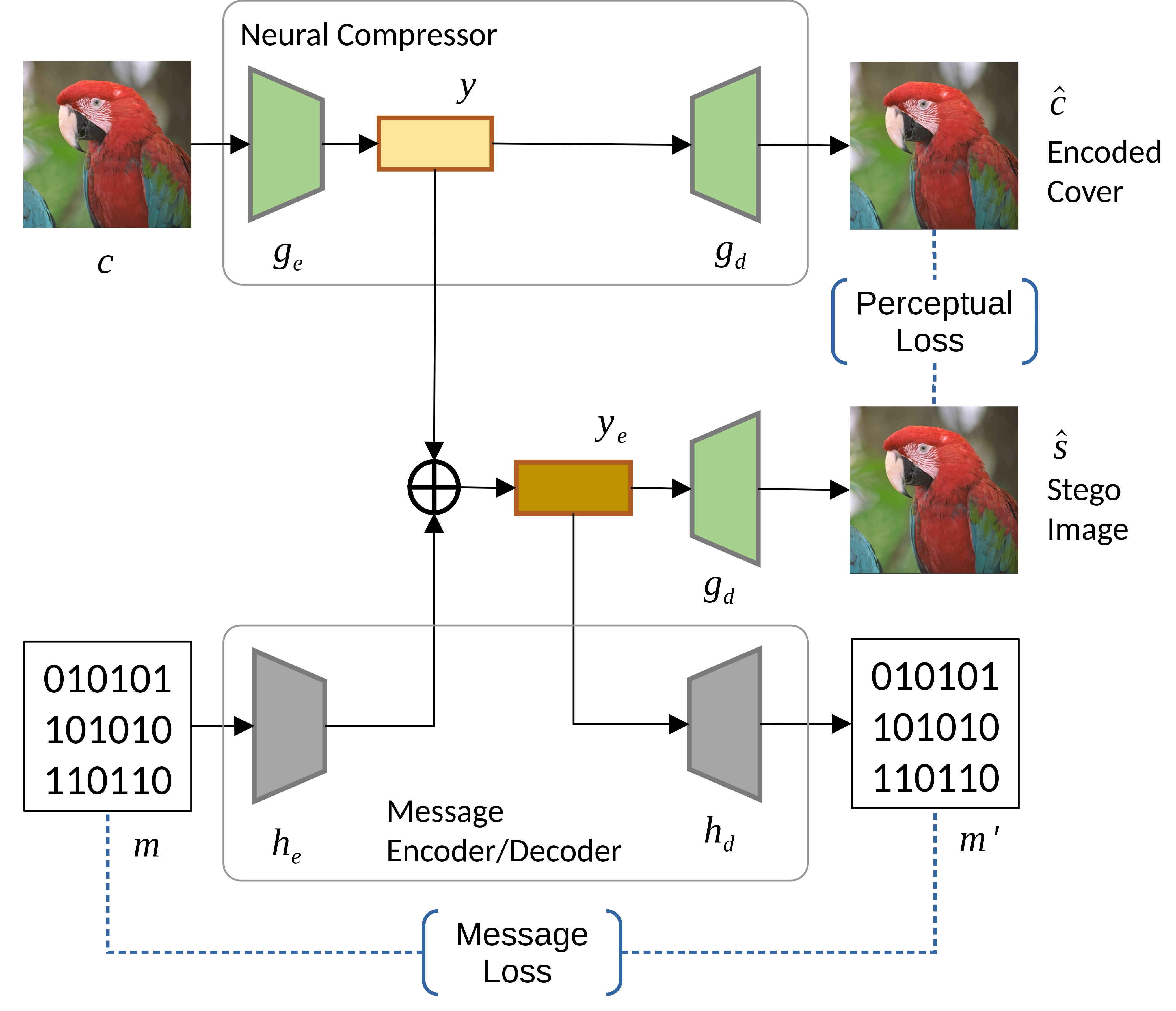}
\caption{The architecture of the proposed neural data hiding framework.}
\label{fig:arch}
\end{figure}

We show our system architecture in Fig. \ref{fig:arch} and describe our techniques as follows.

\subsection{Problem Formulation} \label{sec:Formulation}

\paragraph{Steganography} A message $m\in \{0,1\}^n$ is hidden in a cover image $c$. A neural encoder/decoder pair $g_e$ and $g_d$ are used to obtain the compressed latent vector $y=g_e(c)$ and the encoded cover image $\hat{c}=g_d(y)$. A message encoder $h_e$ is trained to transform the message to the same dimension as $y$. The embedded latent vector $y_e$ is obtained as follows:

\begin{equation}
y_e=y\oplus h_e(m) \label{eq:embed}
\end{equation}

where $\oplus$ is element-wise addition. The embedded latent vector $y_e$ is then entropy-coded and transmitted. The stego image $\hat{s}=g_d(y_e)$ can be decoded by anyone with access to the publicly available $g_e$ and $g_d$. However, only the specified receiver with a trained message decoder $h_d$ can extract the secret message by:

\begin{equation}
m'=h_d(y_e) \label{eq:decode}
\end{equation}


\paragraph{Watermarking} The noised stego image $\dot{s}=u(\hat{s})$ is derived using an attacker $u$ to simulate noise attacks. The final noised stego image $\hat{\dot{s}}$ needs to be re-compressed using the same neural encoder/decoder pair. The noised embedded latent vector $\dot{y_e}=g_e(\dot{s})$ is then used to extract secrets with $\dot{m}'=h_d(\dot{y_e})$.

\subsection{Network Architecture}

We design our message encoder $h_e$ and decoder $h_d$ to match neural codecs \cite{balle2018variational} and \cite{minnen2018joint}'s latent representations. We detail the encoder/decoder structures in Table \ref{tab:my-table}. Different neural codecs may have varying latent space dimensions, but the general rules are: 1) Message encoder: Use a linear layer and a CBNL (convolution, batch normalization, and leaky ReLU) layer to expand the $|m|$-bit message to match the compressed latents' dimensions. 2) Message decoder: Use Conv/ReLU layers with downsampling and then flatten and use linear regression to obtain the $|m|$-bit message.


\begin{table}[!ht]
\centering
\caption{Network architecture}
\label{tab:my-table}
\resizebox{0.9\columnwidth}{!}{
\begin{tabular}{lrrrrrr}
\hline
\hline
\multicolumn{7}{c}{\textbf{Message Encoder $h_e$}} \\ \hline
Layers & kernel & stride & padding & in & out & channels \\ \hline
Linear &  &  &  & $|m|$ & 64 &  \\
CBNL & 3 & 1 & 1 & $8\times 8$ & $8\times 8$ & 320 \\ \hline
\hline
\multicolumn{7}{c}{\textbf{Message Decoder $h_d$}} \\ \hline
Layers & kernel & stride & padding & in & out & channels \\ \hline
Conv/ReLU & 3 & 2 & 1 & $8\times 8$ & $4\times 4$ & 320 \\ 
Conv/ReLU & 3 & 2 & 1 & $4\times 4$ & $2\times 2$ & 320 \\ 
Conv/ReLU & 3 & 2 & 1 & $2\times 2$ & $1\times 1$ & 320 \\ 
Flatten &  &  &  &  & & \\ 
Linear &  &  &  & 320 & 512 &  \\
Linear &  &  &  & 512 & $|m|$ &  \\
\hline
\end{tabular} \label{tab:net-arch}
}
\end{table}

\subsection{Loss Functions}

We define our loss function as a combination of perceptual loss $\mathcal{L}_P$ and message loss $\mathcal{L}_M$:

\begin{equation}
\mathcal{L} = \mathcal{L}_P + \alpha \mathcal{L}_M,
\end{equation}

where $\alpha$ is a hyper-parameter used to control the relative weight of the two losses.

The perceptual loss is measured by the DNN-based perceptual loss LPIPS \cite{zhang2018perceptual}:

\begin{equation}
\mathcal{L}_P = \text{LPIPS}(\hat{c},\hat{s}).
\end{equation}

We avoid using MSE (mean square error) to minimize image distortion because we observed that the LPIPS metric significantly impacts our fixed neural codec more than a self-trained image encoder/decoder.  


For measuring the decoded message error, we use binary cross-entropy as the loss function:

\begin{equation}
\mathcal{L}_M = \text{BCE}(m,m') + \beta \text{BCE}(m,\dot{m}'),
\end{equation}

where $\dot{m}'$ is the decoded message from the noised latent $\dot{y_e}$. The loss function is weighted by hyper-parameters $\alpha=1.5$ and $\beta=1.0$ in our experiments.

\subsection{Noise Attacks}

For the watermarking scenario, we defined four types of common noise attacks as Cropout, Dropout, Gaussian noise, and JPEG compression. We randomly generate the noise parameters during training. 



\section{Experimental Results} \label{sec:experiment}

We implemented our works on neural codecs hyper \cite{balle2018variational} and mbt \cite{minnen2018joint} from CompressAI. We denote our data hiding methods as ``Ours-hyper'' and ``Ours-mbt,'' respectively. To ensure high visual quality after encoding, we set the highest coding quality as 8 in both codecs.

For training, we randomly selected 12,000 and 1,200 images from the COCO dataset \cite{russakovsky2015imagenet} as the training and validation set, respectively. We resized the cover images to $128 \times 128$ and randomly embedded 32-bit binary messages during training. 

We trained our model using the PyTorch built-in Adam optimizer with a learning rate of 0.001 and a batch size of 32. We trained our model for 160 epochs. 
We compared our model with HiDDeN\footnote{https://github.com/ando-khachatryan/HiDDeN}, DeepStega, UDH\footnote{https://github.com/ChaoningZhang/Universal-Deep-Hiding}, and StegaStamp\footnote{https://github.com/tancik/StegaStamp}. Unlike other DNN-based methods that calculate distortion between the cover $c$ and stego image $s$, we measured the distortion between $\hat{c}$ and $\hat{s}$, as described in Section \ref{sec:Formulation}.

\subsection{Steganography Secrecy} \label{sec:secrecy}

Quantitatively, we present image quality metrics in PSNR, SSIM, 
MAE (mean absolute error), and bit error rate, as shown in Table \ref{tab:stego-distortion}. While it is well-known that DNN-based methods cannot achieve zero bit error rates, there are established techniques, such as BCH codes \cite{bose1960class} and learning-based channel noise modeling \cite{choi2019neural}, to mitigate this issue.

Our evaluation accounts for the effect of quantization on latent vectors. As both the hyper and mbt neural codecs use unit scalar quantization, the modified latent coefficients can still withstand the quantization operation. Table \ref{tab:stego-distortion} indicates that our proposed methods have less perceptual distortion than others. The superior stego image quality of the hyper codec stems from its lower coding efficiency than the mbt codec, allowing more room for data hiding in the latent space. On the other hand, the mbt codec has more densely compressed latents, which results in more significant quality degradation from the quantization operation.

\begin{table}[!ht]
\caption{Quality metrics vs. bit error rate comparison}
\label{tab:stego-distortion}
\centering
\resizebox{0.9\columnwidth}{!}{
\begin{tabular}{lrrrr}
\hline
\hline
\multicolumn{5}{c}{\textbf{Kodak}\cite{kodakcd}} \\ \hline
\textbf{Method} & \textbf{PSNR$\uparrow$} & \textbf{SSIM$\uparrow$} & \textbf{MAE}$\downarrow$ & \textbf{Error} \\ \hline
Ours-hyper \cite{balle2018variational} & \First 43.44 & \First 0.9942 & \First 1.02 & 0.00000 \\
Ours-mbt \cite{minnen2018joint} & \Second 40.48 & \Second 0.9881 & \Second 1.65 & 0.00000 \\
HiDDeN \cite{zhu2018hidden} & 39.61 & 0.9813 & 1.91 & 0.00000 \\
DeepStega$^{\mathrm{a}}$ \cite{baluja2017hiding} & 36.51 & 0.9374 & 2.81 & 0.01564 \\ 
UDH$^{\mathrm{a}}$ \cite{zhang2020udh} & 37.88 & 0.9184 & 2.63 & 0.02131 \\
StegaStamp$^{\mathrm{b}}$ \cite{tancik2020stegastamp} & 31.60 & 0.9430 & 4.54 & 0.00500 \\ \hline
\hline
\multicolumn{5}{c}{\textbf{DIV2K}\cite{Agustsson_2017_CVPR_Workshops}} \\ \hline
\textbf{Method} & \textbf{PSNR$\uparrow$} & \textbf{SSIM$\uparrow$} & \textbf{MAE}$\downarrow$ & \textbf{Error} \\ \hline
Ours-hyper & \First 41.67 & \First 0.9945 & \First 1.36 & 0.00000 \\
Ours-mbt & \Second 38.42 & \Second 0.9847 & \Second 2.25 & $^{\mathrm{c}}$0.00094 \\
HiDDeN & 37.59 & 0.9733 & 2.44 & 0.00125 \\
DeepStega & 34.72 & 0.9283 & 3.58 & 0.01839 \\ 
UDH & 38.35 & 0.9414 & 2.51 & 0.02475 \\ 
StegaStamp & 30.50 & 0.9451 & 5.38 & 0.00890 \\ \hline
\hline
\multicolumn{5}{c}{\textbf{CelebA}\cite{liu2015deep}} \\ \hline
\textbf{Method} & \textbf{PSNR$\uparrow$} & \textbf{SSIM$\uparrow$} & \textbf{MAE}$\downarrow$ & \textbf{Error} \\ \hline
Ours-hyper & \First 46.08 & \First 0.9962 & \First 0.75 & 0.00000 \\
Ours-mbt  & \Second 40.56 & 0.9768 & 2.25 & 0.00000 \\
HiDDeN & 39.28 & \Second 0.9806 & \Second 1.85 & 0.00000 \\
DeepStega & 38.27 & 0.9410 & 2.34 & 0.01822 \\ 
UDH & 38.27 & 0.9147 & 2.54 & 0.01502 \\ 
StegaStamp & 35.40 & 0.9586 & 2.67 & 0.00460 \\ \hline
\multicolumn{5}{l}{$^{\mathrm{a}}$Use the change of MSB to report bit error} \\
\multicolumn{5}{l}{$^{\mathrm{b}}$Image size $400\times 400$, embed 100 bits} \\
\multicolumn{5}{l}{$^{\mathrm{c}}$Bit errors due to quantization} \\
\end{tabular}
}
\end{table}

\begin{figure}[!t]
\centering
\includegraphics[width=1\columnwidth]{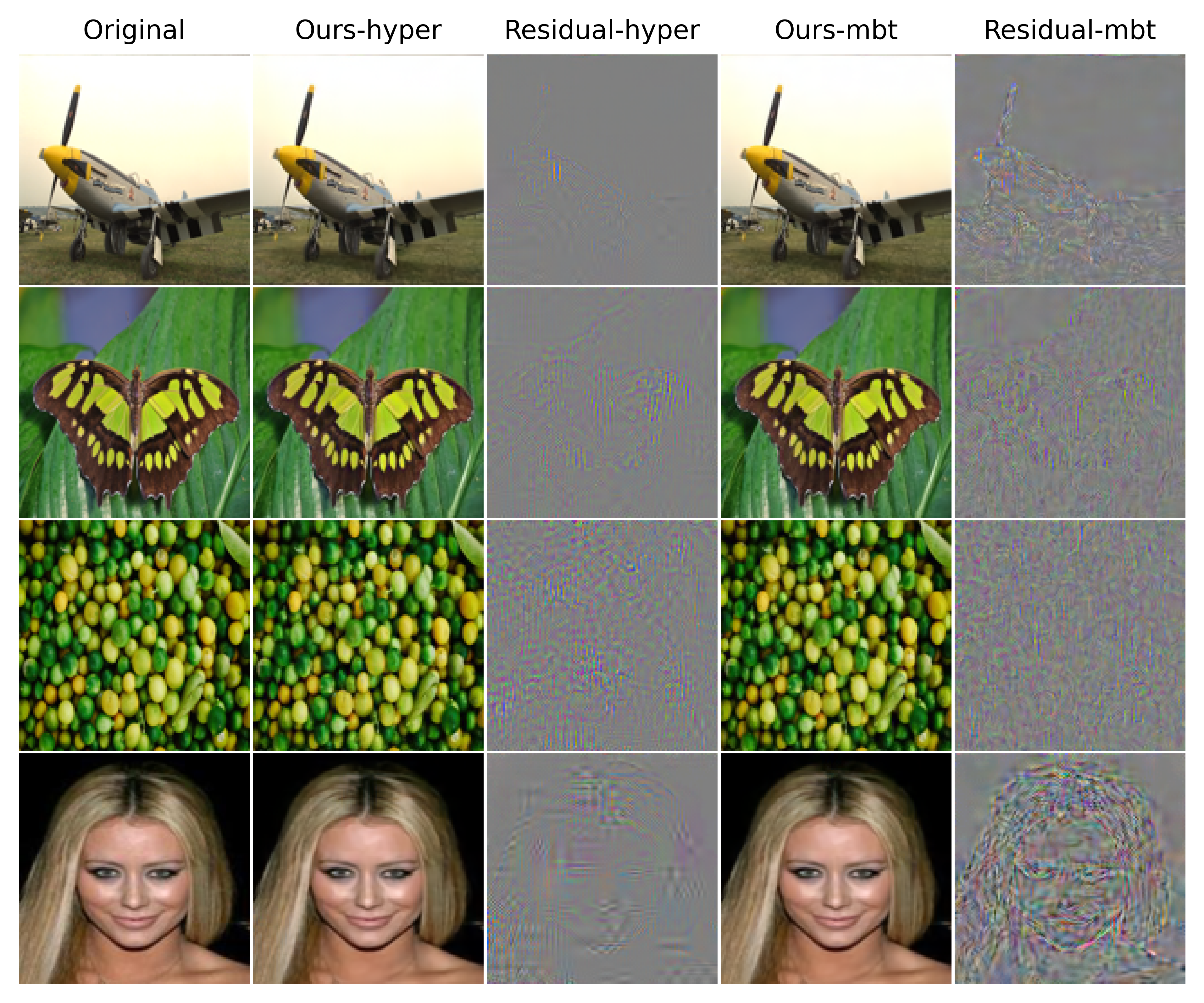} 
\caption{Comparison of stego image quality and residual. Please zoom in to observe the modified pixel locations.}
\label{fig:embed-quality}
\end{figure}

\begin{figure}[!t]
\centering
\includegraphics[width=1\columnwidth]{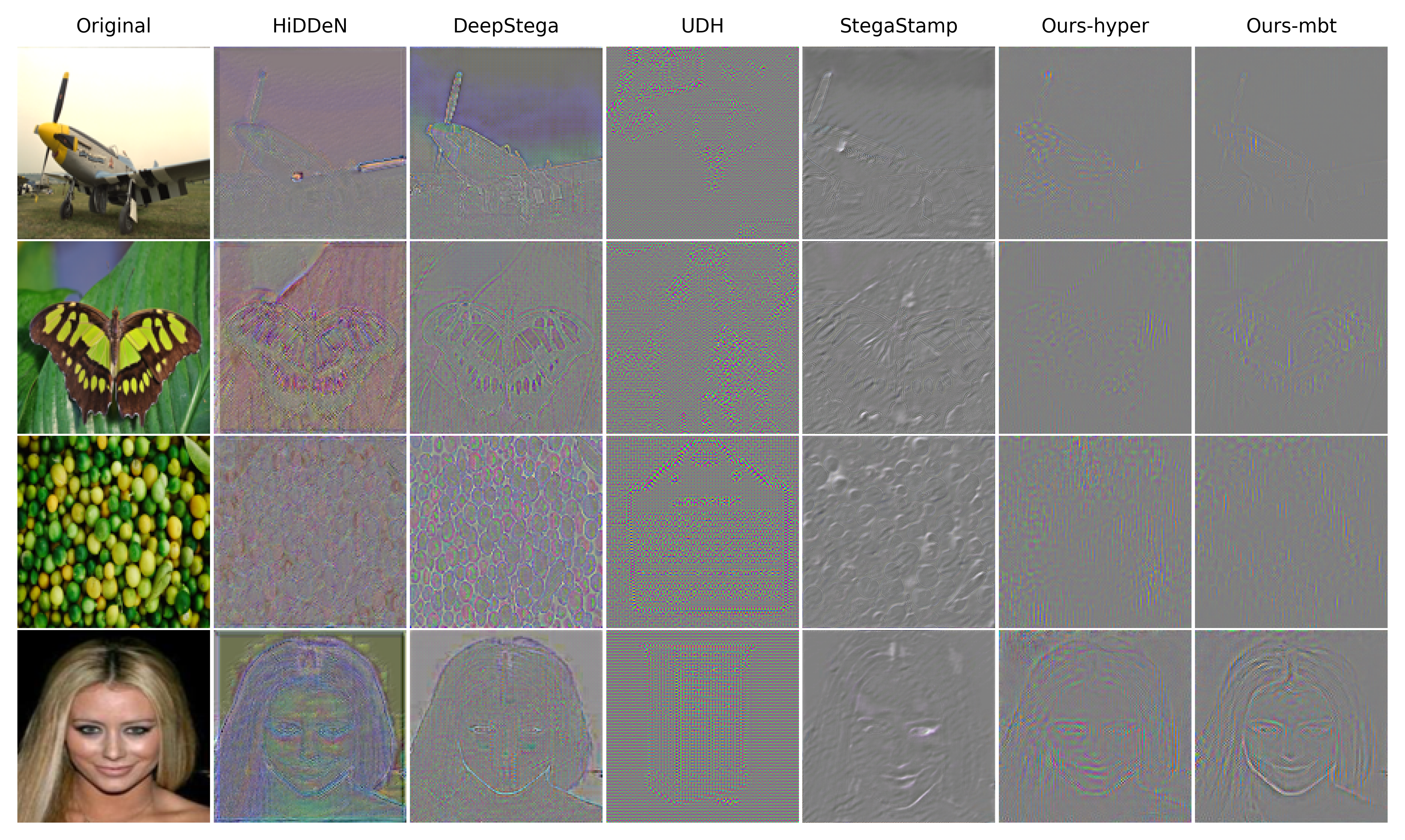}  \\
\caption{The stego image residual comparison of different data hiding methods. Please zoom in to observe the modified pixel locations.} 
\label{fig:embed-residual}
\end{figure}

Qualitatively, we present the cover images and the resulting stego images in Fig. \ref{fig:embed-quality} and compare our methods' stego image residual with other DNN-based methods in Fig. \ref{fig:embed-residual}. 

\paragraph{Pixel modification} Modern DNN-based data hiding methods add perturbations to the low-level feature space to extract messages from the spatial domain with robustness. As a result, these methods modify all the low-level pixels of the cover image, as shown in Fig. \ref{fig:embed-residual}. Our neural data hiding method learns to modify the compressed latents, so the pixel modifications are placed in high-level image features, as shown in Fig. \ref{fig:embed-quality}. Overall, our proposed method can generalize well on different neural codecs and has a less perceptual impact.

Although in our scenario, the receiver does not have access to the encoded cover image $\hat{c}$, we list the LPIPS \cite{zhang2018perceptual} metrics between $c$ and $\hat{s}$ in Table \ref{tab:lpips-stego} and compare them with those of other methods. Our LPIPS metrics remain close to those of the HiDDeN method and are superior the other methods. The LPIPS is a learned perceptual metric based on a pre-trained DNN, which can be thought of as how effectively the stego image can be used as a proxy for the original cover image. Therefore, we believe the generated stego image $\hat{s}$ has not lost its general utility.

\begin{table}[!ht]
\caption{LPIPS comparison of stego images on DIV2K}
\label{tab:lpips-stego}
\centering
\resizebox{0.8\columnwidth}{!}{
\begin{tabular}{lrrr}
\hline
\textbf{Method} & \textbf{$\text{LPIPS}(\hat{c},\hat{s})$} & \textbf{$\text{LPIPS}(c,\hat{s})$} & \textbf{$\text{LPIPS}(c,s)$} \\ \hline
Ours-hyper & 0.00064 & 0.00392 & \multicolumn{1}{c}{-} \\
Ours-mbt & 0.00485 &  0.00599 & \multicolumn{1}{c}{-} \\
HiDDeN & \multicolumn{2}{c}{-} & 0.00375  \\
DeepStega & \multicolumn{2}{c}{-} & 0.07718 \\ 
UDH & \multicolumn{2}{c}{-} & 0.04261  \\ 
StegaStamp & \multicolumn{2}{c}{-} & 0.08039 \\ \hline
\end{tabular}
}
\end{table}

\paragraph{Steganalysis} We measured the ability of our model to resist steganalysis using publicly available steganalysis tools, including traditional statistical methods \cite{boehm2014stegexpose} and new DL-based approaches \cite{lerch2016unsupervised} \cite{boroumand2018deep}. To assess our model's anti-steganalysis ability on the DIV2K dataset, we used the steganalysis tool StegExpose \cite{boehm2014stegexpose}.

\begin{figure}[!t]
\centering
\includegraphics[width=0.9\columnwidth]{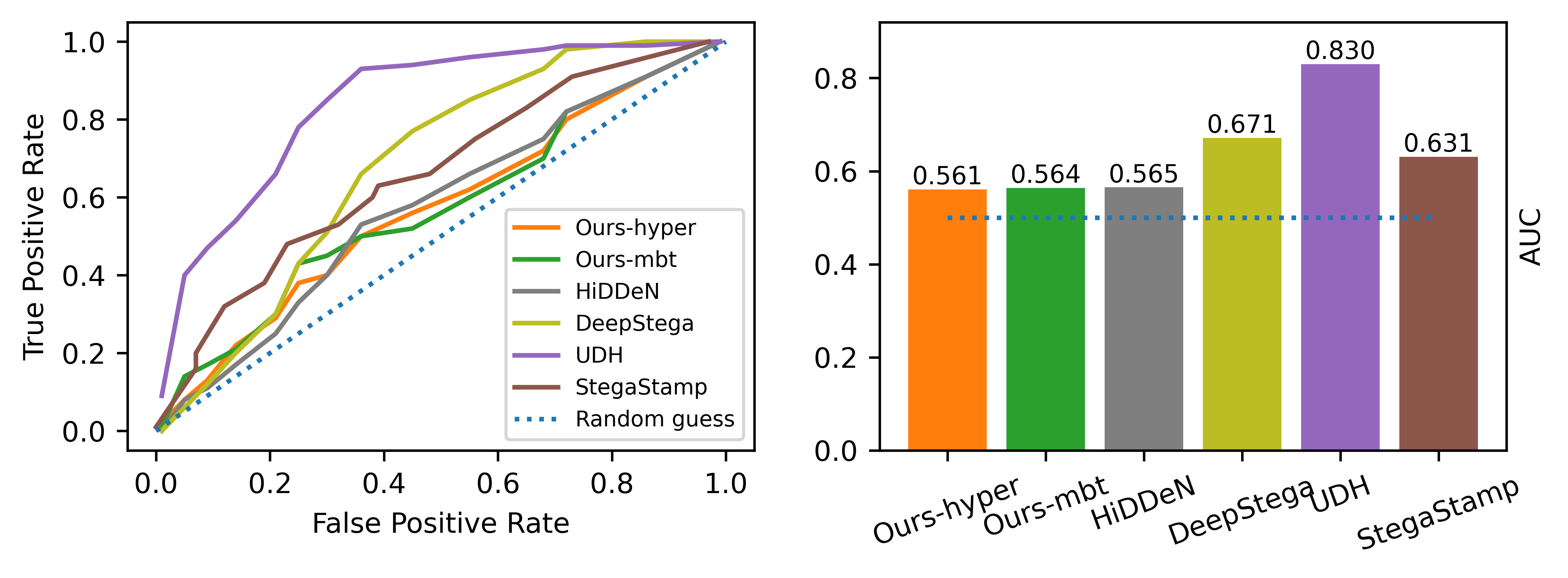}
\caption{The ROC curve of StegExpose classifier.}
\label{fig:steganalysis}
\end{figure}

We varied the detection thresholds as input to StegExpose and plotted the ROC (receiver operating characteristic) curve. We then calculated the AUC (area under the curve) to indicate the classification effectiveness. Ideally, the AUC should be close to 0.5, indicating that the classifier performs no better than random guessing. Figure \ref{fig:steganalysis} shows the ROC curve and the AUC of the compared methods. Our proposed Ours-hyper method achieved slightly better secrecy than HiDDeN, with an AUC of 0.561.

\subsection{Watermark Robustness}

\begin{figure}[!ht]
\centering
\includegraphics[width=\columnwidth]{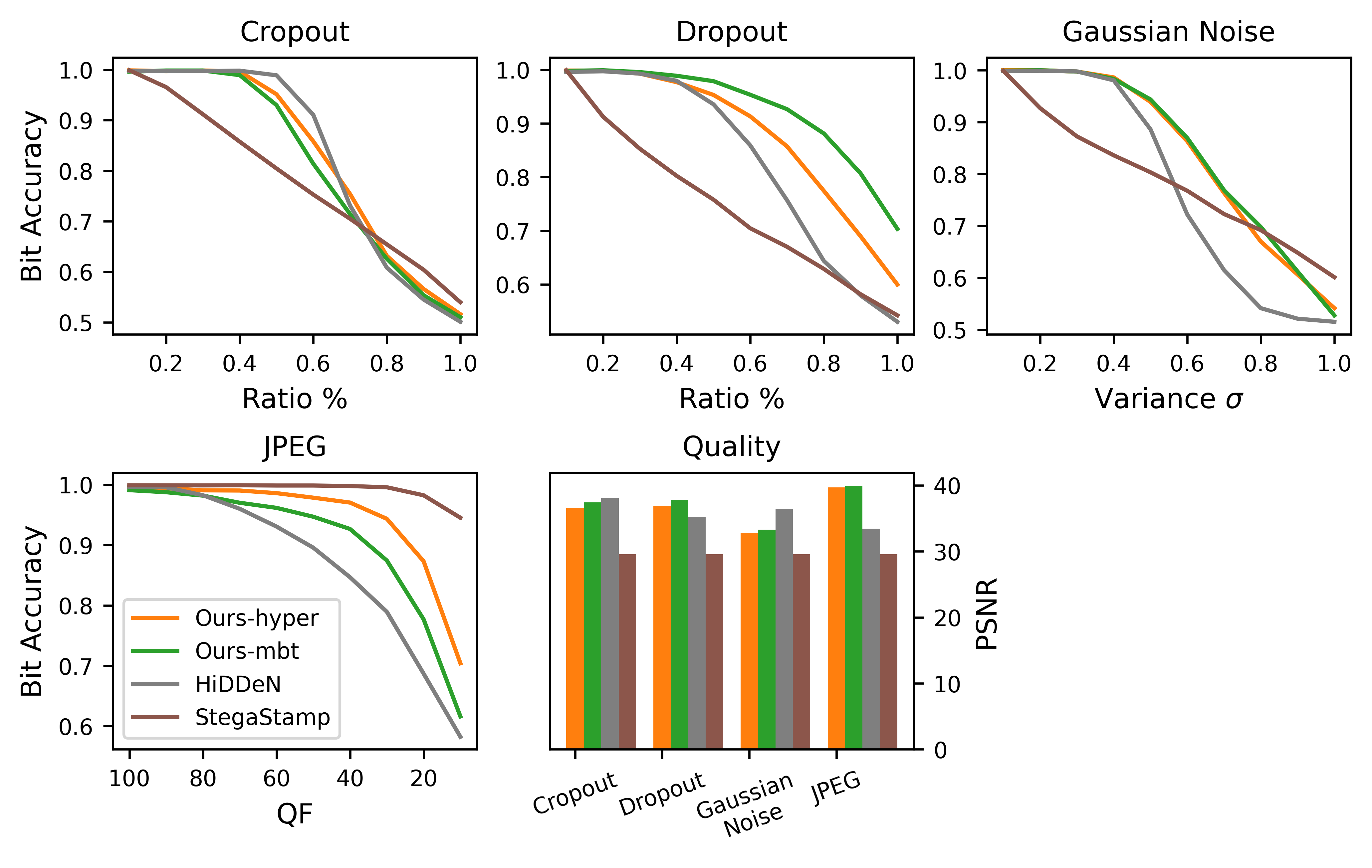}  \\
\caption{Watermark robustness against selected noise attacks, evaluated on DIV2K.} 
\label{fig:noise-robust}
\end{figure}

We evaluated the robustness of our method against trained noise attacks on the DIV2K dataset, as shown in Fig. \ref{fig:noise-robust}. We varied the attack strength by increasing the noise parameter, which degrades image quality along the horizontal axis. In the last chart of Fig. \ref{fig:noise-robust}, we observe that the overall PSNR is lower than the numbers we reported in Table \ref{tab:stego-distortion}, which is the trade-off we make for achieving robustness.

Our neural data hiding method performs equivalently to HiDDeN in the cropout attack, as the cropped-out pixels do not provide any information to the message extractor. However, for attacks such as dropout, Gaussian noise, and JPEG compression, the proposed method demonstrates superior robustness to spatial domain extraction methods like HiDDeN and StegaStamp. We believe this is due to modifying of the compressed latent space, which retains more information in the high-level features without being impacted.

\subsection{Embedding Performance}

We report the message embedding/extraction time in Table \ref{tab:speed}, measured on an Intel i7-9700K workstation with an Nvidia GTX 3090 GPU. As expected, our method embeds messages 50 times faster than other spatial domain approaches. 

\begin{table}[!htb]
\centering
\begin{minipage}[t]{.45\linewidth}
    \caption{Embed/extract time on DIV2K, in seconds.}
    \label{tab:speed}
    \centering
    \resizebox{\textwidth}{!}{
\begin{tabular}{lrr}
\hline
\textbf{Method} & \textbf{Embed} & \textbf{Extract}  \\ \hline
Ours-hyper & 0.00021 & 0.00051 \\
Ours-mbt & 0.00020 & 0.00049  \\
HiDDeN & 0.01040 & 0.00074 \\
DeepStega & 0.01108 & 0.00071 \\ 
UDH & 0.01091 & 0.00070 \\ 
StegaStamp & 0.05893 & 0.03412 \\ \hline
\end{tabular}
    }
\end{minipage}%
\quad
\begin{minipage}[t]{.5\linewidth}
    \caption{Message embedding size overhead, in percentage.}
    \label{tab:overhead}
    \centering
    \resizebox{\textwidth}{!}{
\begin{tabular}{lrrr}
\hline
\textbf{Method} & \textbf{Kodak} & \textbf{DIV2K} & \textbf{CelebA} \\ \hline
Ours-hyper & 15.62\% & 11.95\% & 25.61\% \\
Ours-mbt & 20.51\% & 14.38\% & 31.18\% \\ 
HiDDeN$^\mathrm{a}$ & 4.36\% & 5.03\% & 9.71\% \\ \hline
\multicolumn{4}{l}{$^{\mathrm{a}}$In compressed PNG} \\
\end{tabular}
    }
\end{minipage}
\end{table}

\subsection{Neural Codecs}

We point out some key considerations to be taken into account when building applications on neural compressed latent representations:

First, the continuous latents will be quantized before entropy coding. Although in our case, the modified latent coefficients can still withstand unit quantization being used, it is worth further investigating how different quantization operations can affect the message embedding process.

Second, the impact on coding efficiency. The neural compressor learns a compact representation and a near-optimal probability estimation end-to-end. Adding perturbations to latent coefficients inevitably breaks the compressor's learned optimal coding strategy. 
The overheads shown in Table \ref{tab:overhead} range from 11\% to 31\% depending on the dataset and the underlying codecs used.

We did not jointly optimize the image codec with the message encoder/decoder because we aimed to prove the concept of neural data hiding in a standard neural compressor. Third, we observed that a more coding efficient neural compressor leads to more quality degradation from quantization and more size overhead after embedding. 

\subsection{Perceptual Loss Study} \label{sec:ablation}

The success of message extraction is highly dependent on the design of the message encoder/decoder when jointly trained with a standard codec that minimizes image distortion. Experimental results show that the widely used MSE loss performs poorly. Interestingly, we observed that the LPIPS metric significantly impacts our fixed neural codec more than a self-trained image encoder/decoder. This observation could be due to the high compactness of our neural codec, which is more strongly connected to certain perceptual features in the latent space.

\section{Conclusion}

In this work, we proposed a novel end-to-end framework for image data hiding that embeds secrets in the latent representations of a neural compressor. Our approach is generic and can be used with different neural compressors. We demonstrated its superior image secrecy and competitive watermarking robustness while significantly accelerating the embedding speed. 
Future researches could focus on designing more efficient neural compressors that reserve space for data hiding in their latent vectors.


\bibliographystyle{IEEEtran}
\bibliography{all_refs}


\end{document}